\documentclass[a4paper, epsf]{spie}  
\usepackage[]{graphicx}

\def\farcm@mss{\ensuremath{.\mkern-4mu{}^{\prime}}}%
\def\farcs@mss{\ensuremath{.\!\!{}^{\prime\prime}}}%

\def\gee{ \, \lower 1mm\hbox{$\,{\buildrel > \over{\scriptstyle\scriptstyle\sim}
}\displaystyle \,$}}
\def\lee{ \, \lower 1mm\hbox{$\,{\buildrel < \over{\scriptstyle\scriptstyle\sim}
}\displaystyle \,$}}
\def\|{\partial}

\def\varkappa {{\scriptstyle\partial}\! e}

\title{Static Pressure of Hot Gas: \\ Its Effect on the Gas Disks of Galaxies}
\author{Anatoly V. Zasov and Alexander V. Khoperskov \\[12pt]
Moscow State University, RUSSIA\\[12pt]
Volgograd State University, RUSSIA\\[12pt]
}

\authorinfo{\textit{Journal-ref}: \textbf{Astronomy Letters, 2008, Vol. 34, No. 11, pp. 739--744}. \\
\vspace{5mm} A.Zasov: E-mail: zasov@sai.msu.ru; A.Khoperskov:
E-mail: khoperskov@rambler.ru}

\let\b=\baselineskip
\begin{document}
\maketitle

\begin{abstract}
The static pressure of the hot gas that fills clusters and groups
of galaxies can affect significantly the volume density and
thickness of the gas disks in galaxies. In combination with the
dynamic pressure, the static pressure allows several observed
peculiarities of spiral galaxies surrounded by a hot medium to be
explained.
\end{abstract}

\keywords{gas in clusters of galaxies, star formation in galaxies,
interaction of galaxies with ambient medium}

\section{INTRODUCTION}

\b=1.45\b

The interstellar gas in galactic disks is nonuniform in density.
However, since the denser regions are generally colder, the pressure
fluctuations are considerably smaller than the density fluctuations
and we can talk about characteristic equilibrium values of the
pressures $P$ at a given galactocentric distance $r$.

In the solar neighborhood, the pressure of the interstellar medium
(normalized to the Boltzmann constant $k$) due primarily to the
turbulent motions of gas is $\simeq\ 2\cdot 10^{4}$K$\cdot
$cm$\!^{-3}$ (Cox 2005), i.e., $\log P/k\approx 4.3$. Since the
pressure of the gas is determined primarily by its mean density, it
decreases with galactocentric distance. According to the
calculations of the equilibrium pressure of the gas disk in the
plane of the stellar disk for self-consistent models of several
nearby spiral galaxies, including our Galaxy, (Kasparova and Zasov
2008), the logarithmof the gas pressure $P/k$ in the chosen system
of units is $4-5$ at $r=(0.2-0.3)\, R_{25}$ and $3.5-4$ at $r =
(0.7-0.8)\,R_{25}$, where $ R_{25}$ is the photometric radius of the
galaxy. At even larger galactocentric distances, the gas disk
expands rapidly and the pressure falls sharply. The gas pressure is
particularly low in low-brightness galaxies, where the gas volume
density is an order of magnitude lower than that in ordinary spiral
galaxies. Nevertheless the star formation, although slow, still
holds even there.

If a galaxy is in a cluster, then its interstellar medium is
affected by a hot intergalactic gas. For a rapidly moving galaxy and
a favorable orientation of its disk with respect to its velocity
vector, the dynamic (ram) pressure of the gas, being proportional to
$n_e V_c^2$, where $n_e$ is the electron number density of the outer
gas and $V_c$ is the relative velocity of the galaxy, should
manifest itself. The ram pressure sweeps up the tenuous gas ($HI$)
from the outer regions of galaxies, produces an asymmetry in the
$HI$ distribution, and reduces the radius of the region occupied by
gas in HI-deficient galaxies (see, e.g., Scodeggio and Gavazzi 1993;
Cayatte et al. 1994; and references therein). It is much more
difficult to sweep up the gas from the inner regions of massive
galaxies, where the stellar disk is denser and the potential well
produced by the gravitational field of the galaxy is much deeper.

However, the pressure of the surrounding gas at a fairly high
temperature can be significant even for a slowly moving galaxy in
the hot gas medium. Since the hot gas fills the entire cluster, the
speed of sound $c\approx (P/\rho)^{1/2}$ for it is close to the
root-mean-square velocity of the galaxies. Therefore, the static
pressure of the intergalactic gas $P=2 n_ekT$ due to the electron
and ion components is close to the mean dynamic pressure. In
contrast to the latter, the static pressure acts independently on
the galaxy's velocity relative to the medium or the orientation of
its disk; it manifests itself at all galactocentric distances.

The available temperature and density estimates for the
intergalactic medium in systems of galaxies show that the static
pressure $P$ of the hot gas is generally comparable to the estimated
pressure of the interstellar gas in galactic disks and, hence, can
affect significantly the evolution of the gas inside the galaxy.
Indeed, the characteristic temperatures of the hot intergalactic gas
range from several keV (more than $10^7$~K) in small clusters to 10
keV (more than $10^8$K) in such clusters as Coma (see, e.g., Arnaud
and Evrard 1999; Finoguenov et al. 2001). In rich clusters the gas
density decreases with distance from the cluster center according to
a law that is usually approximated by the formula
\begin{equation}\label{Eq-concentr}
    n(R) = n_0 \, \left[ 1 + \left( \frac{R}{R_c} \right)^2 \right]^{-3\beta/2} \,,
\end{equation}
where the parameter $\beta \simeq 0.4 - 0.7$ depends on the internal
structure of the cluster. The core radius $R_c$ is usually several
tens of kpc for small clusters and several hundred kpc for the
largest systems (see, e.g., Arnaud and Evrard 1999). At a distance
of $(1- 2)\cdot R_c$, the electron number density is $n_e\simeq
10^{-2} - 10^{-3}$cm$^{-3}$. Indeed, in such rich clusters as Coma,
A 1795, and A 3112, $\log~P/k\gee 5$ within several hundred kpc from
the center (Nevalainen et al. 2003). In the Virgo cluster, the
particle density at a distance of 100--200 kpc from the center at a
temperature $T\approx 3$ keV is $10^{-3}$cm$\!^{-3}$  (Nulsen and
Bohringer 1995), which corresponds to $\log P/k \gee 4.5$. Even at
$n_e\sim 10^{-4}$, which is more characteristic of the cluster as a
whole, we obtain $\log~P/k \gee 3.5-4$ for such clusters as Virgo.
This value exceeds the expected pressure of the interstellar gas in
the outer regions of spiral galaxies. In small clusters containing
an X-ray gas, the pressure is of the same order of magnitude as that
in large ones: $kT \approx 1.5$ keV, $n_e\approx
10^{-3}$cm${\!}^{-3}$ (Dahlem and Thiering 2000).

The same reasoning is also applicable to galaxies in groups, if the
latter are filled with an X-ray-emitting gas. In this case, the
slightly lower temperature of the hot gas ($\sim $ 1 keV) is
compensated for by the higher particle number density. There exists
an $HI$ deficit in galaxies of such groups in spite of their low
relative velocity dispersion, and this deficit is more significant
than that in X-ray dim groups (Sengupta et al., 2007).

Thus, in systems of various scales the pressure on the gas disk can
exceed the internal pressure of the interstellar gas that would
exist in the absence of an external action. As a result, the
pressure of the interstellar gas in galaxies, particularly in the
central regions of clusters, should be, on average, higher and the
gas disk thickness $2h$ should be smaller than those for galaxies
that undergo no external pressure. The lower the gas density in a
galaxy, the stronger the effect of the static pressure. On a
qualitative level, this question was discussed previously (Zasov
1987).

\section{THE PRESSURE IN THE GAS DISK
IN THE PRESENCE OF AN INTERGALACTIC MEDIUM }

Consider the increase in gas pressure in the galactic plane
quantitatively, in terms of simple equilibrium models. Let us
write the hydrostatic equilibrium equation for the gas in the
gravitational field of the stellar disk,
\begin{equation}\label{Eq-hydro-stat-equil}
    \frac{dP}{dz} = - \varrho(z)\,g(z) = -
    \frac{\mu}{{\cal R}}\,\frac{P(z)}{T(z)}\, g(z)
     \,.
\end{equation}
The acceleration $g(z)$ can be expressed in terms of the vertical
oscillation frequency $\Omega_z$: $g(z)=z\,\Omega_z^2
/(1+|z|/\Delta)$, where $\Delta$ is the vertical scale height of
the stellar disk. This formula describes a linear increase in
$g(z)$ at small $z$ with its value reaching a constant at large
heights $|z| \gg \Delta$. If we restrict ourselves to the
approximations $g\propto z$ and $T=T_0=\textrm{const}$, then the
solution $P=P_0\exp(-z^2/2h_0^2)$ follows from
(\ref{Eq-hydro-stat-equil}), where we have  $h_0^2={\cal
R}T_0/\mu\Omega_z^2$, for the vertical scale height $h_0$ of an
isothermal gas disk.

\begin{figure}[!t]
 \vskip 0.\hsize \hskip 0.25\hsize
            \includegraphics[width=0.5\hsize]{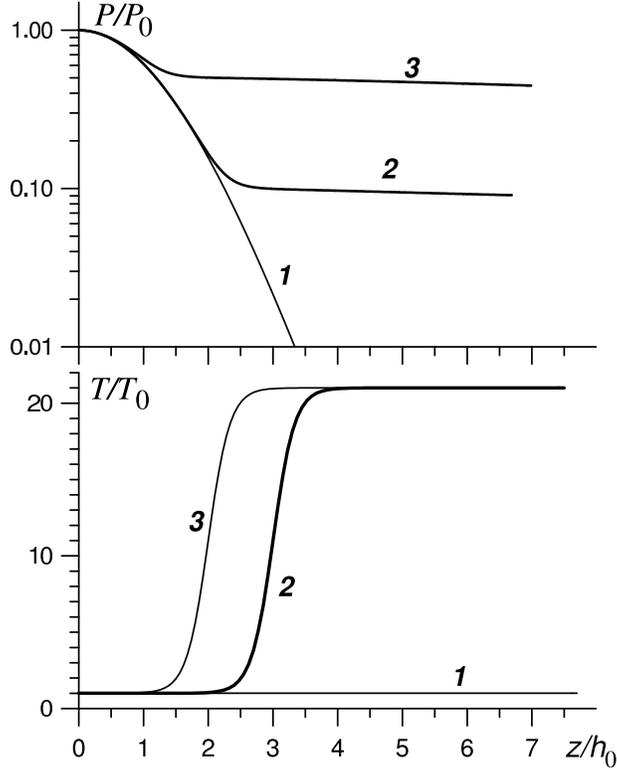} \vskip -0.\hsize

\vskip  -0.05\hsize \hskip 0.\hsize
  \vbox{\hsize=0.99\hsize
  \caption {
 ``Vertical'' profiles of the pressure (a) and temperature (b)
normalized to their values in the $z=0$ plane for the adopted
ratio of the half-thicknesses of the unperturbed gas and stellar
disks $h_0/\Delta=0.2$: \textit{1}, in the absence of an
atmosphere ($T=\textrm{const}$); \textit{2} and \textit{3}, in the
presence of an atmosphere. Curve \textit{3} corresponds to a
higher pressure of the atmosphere $P_a$. }
 \label{Fig-p(z)} }\vskip 0.0\hsize
\end{figure} 

We will consider the hot gas that surrounds the disk as an
atmosphere with a certain pressure $P_a$ and temperature $T_a$.
Let the temperature be $T=T_0=\textrm{const}$ in the disk and be
also constant at larger heights in the atmosphere, with $T(z\gg
h_0)=T_a \gg  T_0$, so that the temperature passes from $T_0$ to
$T_a$ in some zone at $z>h_0$. Figure~\ref{Fig-p(z)} illustrates
the pressure profiles along $z$ obtained by numerically
integrating (\ref{Eq-hydro-stat-equil}) for various dependences of
the temperature $T(z)$ on the vertical coordinate. As we see from
the figure, for a fairly sharp increase in temperature at a height
$z\sim (2-3) \, h_0$, the pressure profile ceases to decline with
height, reaching a plateau $P=P_a$.

The mean gas pressure in a galactic disk can be represented as
 \begin{equation}\label{Eq-balance-simple}
    {\langle {P}\rangle} = {P_a} + \langle
    \varrho\rangle\, g h = {P_a} + \langle
    \varrho\rangle\, \Omega_z^2 h^2 \,,
\end{equation}
where $h$ is the gas scale height in the presence of the atmosphere,
the ``vertical'' acceleration within $h$ due to the stellar disk
gravity is assumed to be $g=\Omega_z^2 h$, and the angular brackets
$\langle ... \rangle$ denote an average over the $z$ coordinate. Let
us denote $\langle \varrho\rangle = k_1\varrho_0$ and $\langle
{P}\rangle = k_2{\cal P}_0$, where the subscript ``$0$'' refers to
the quantities in the $z=0$ disk plane and the coefficients
$k_{1,2}$ are determined by the pattern of the vertical density and
pressure profiles ($0<k_{1,2}<1$).

Let the gas volume density be $\varrho_1$ in the absence of an
external atmosphere and $\varrho_2$ in its presence and the
corresponding values for the characteristic gas disk
half-thicknesses be $H$ and $h$. Obviously, $H=h=h_0$ and $h < H$
for an isothermal model without and with an atmosphere,
respectively.

\begin{figure}[!t]
 \vskip 0.\hsize \hskip 0.0\hsize
            \includegraphics[angle=-90,width=0.999\hsize]{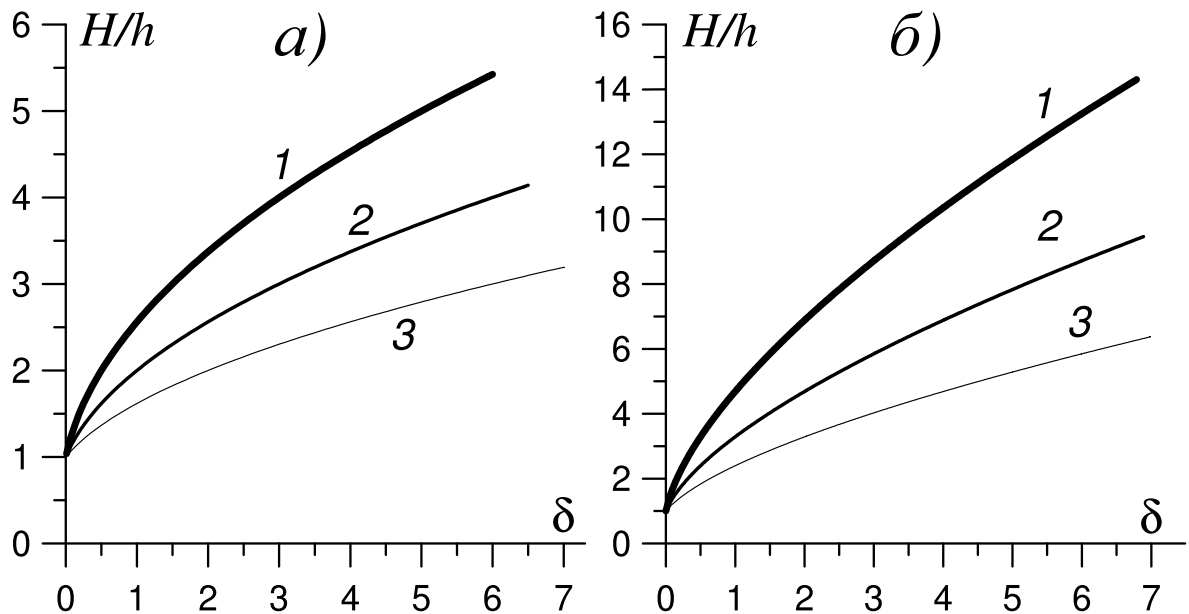}

\vskip  0.05\hsize \hskip 0.0\hsize
  \vbox{\hsize=0.9900\hsize
  \caption {
Ratios of the characteristic gas disk half-thicknesses in the
absence of an atmosphere ($H$) and in its presence ($h$) versus
ratio $\delta$ of the external pressure $P_a$ to the unperturbed
pressure $P_0$ for various parameters of the vertical gas density
profile: (a) $n=2$ and (b) $n=1.4$; \textit{1}, $k_1=k_2=0.25$;
\textit{2}, $k_1=k_2=0.5$; \textit{3}, $k_1=k_2=1$. The last case
corresponds to a disk with a constant density at $|z|\le h$.
 }\label{Fig-x(delta)} }\vskip 0.0\hsize
\end{figure} 

We will assume that in the regions where the gas was not swept up
from the disk by the ram pressure, its surface density is conserved
during compression and, given the mass conservation in the disk, we
can write

 $$H\varrho_{01} = h\varrho_{02}$$.

Taking into account the last relation and restricting ourselves to a
polytropic law with a polytropic index $n$,  we can write

 \begin{equation}\label{Eq-politrop}
{\cal P}_0 = P_0\, \left(\frac{H}{h}\right)^n \,,
\end{equation}
where ${\cal P}_0$ is the midplane gas pressure in the absence of
the atmosphere.

Substituting (\ref{Eq-politrop}) into (\ref{Eq-balance-simple}), one
can obtain
\begin{equation}\label{Eq-x-delta}
 x^{n-1} + \frac{\delta}{k_1}\, x^n - \frac{k_2}{k_1} = 0 \,,
\end{equation}
where $x=h/H$ and $\delta = P_a / P_0$. For $n=2$, we find for the
ratio of the vertical scale heights in the presence and the absence
of an atmosphere:
\begin{equation}\label{Eq-x(delta)-n=1}
    x = \frac{h}{H} = \frac{1}{2}\left( \sqrt{4\frac{k_2}{\delta}+
    \frac{k_1^2}{\delta^2} } -
\frac{k_1}{\delta} \right) \,.
\end{equation}
For $\delta = 0$ (there is no hot atmosphere), we will set
$k_1=k_2$ to satisfy the condition $x=1$. In the other extreme
case, the asymptotics $x\propto 1/\sqrt{\delta}$ takes place for a
large pressure difference (Fig.~\ref{Fig-x(delta)}a). The models
with a lower value of the index $n$ seem more realistic. As an
example, Fig.~\ref{Fig-x(delta)}b shows the ratios $H/h$ as a
function of $\delta$ for $n=1.4$. We see that the disk becomes
appreciably thinner with decreasing $n$.

Thus, as the external static pressure from the intergalactic hot gas
increases, the thickness of the gas disk inside a galaxy can
decrease significantly; the proportionality coefficient between $H$
and $h$ is determined by the equation of state for the gas and by
the pattern of the vertical density (and pressure) profile, which,
in turn, depends on the chosen thermodynamic model of the gas. For
instance, as follows from Fig.~\ref{Fig-x(delta)}, at an external
pressure equal to the unperturbed pressure in the plane of the disk,
its thickness decreases and, hence, the gas volume density increases
by a factor of $1.5-4$. If, however, the pressure of the atmosphere
$P_a$ exceeds $P_0$, say, by a factor of 4, then the disk thickness
changes by a factor of $2.5-10$; the lower values of the above
quantities refer to a clearly unrealistic model with a uniform
density distribution in disk height. Note, however, that the
presence of a magnetic field in the interstellar medium, whose
pressure is initially comparable to the interstellar gas pressure,
will slightly reduce the gas compression ratio, since in the case of
magnetic flux conservation the magnetic pressure, which counteracts
the compression, increases as $(H/h)^2$. In the limiting case, if
the interstellar gas pressure under strong compression is low
compared to the magnetic field pressure, the field strength is
$B=\sqrt{8\pi P_a}$. Note that the magnetic field can also greatly
reduce the heat conduction at the interface between the two media,
isolating the colder gas inside the galaxy from the hot environment.
This applies not only to the gas disks of galaxies in clusters, but
also to the gas halos of galaxies, which contract, but ``survive'',
despite being surrounded by a hotter medium (see, e.g., Sun et al.
2007; Vikhlinin et al. 2001).

Thus, one might expect the static pressure of the environment in
clusters and groups of galaxies to be capable of increasing the
density of the interstellar medium that was not swept up by the
dynamic pressure of intergalactic gas by a factor of several. As a
result, the gas disks in this case should be, on average, thinner,
while the gas midplane volume density should be higher at a fixed
column density. In turn, this means a significant increase in star
formation rates per unit gas mass and a faster gas exhaustion in the
entire disk.

\section{DISCUSSION}

The static gas pressure is the most universal mechanism of the
influence of the ambient medium on a galaxy, since it acts on a  gas
layer in all cases where the galaxy is surrounded with a hot medium.
Below we discuss some data supporting the assumption that the gas
disk is compressed by an external pressure in many galaxies,
especially if they are located in the inner regions of clusters,
although the arguments given remain indirect. Therefore, in each
specific case, we cannot rule out the action of other factors
either. In particular, hydrodynamic calculations demonstrate that a
strong ram pressure may influence the star formation rate not only
in the periphery, but also in the inner parts of a galaxy
(Kronberger et al., 2008). It is evident that in general case both -
dynamic and static pressures - should be considered as a single
process. Note however that the efficiency of the ram pressure unlike
the static one is proportional to $\rho_{gas} V_c^2$, hence it is
not too effective if the velocity $V_c$ of a galaxy is lower than
the mean velocity dispersion of galaxies in a cluster.

(1) Many spiral galaxies in clusters, including $HI$-deficient
ones, are actually distinguished by high star formation rates per
unit gas mass, i.e., by a short gas exhaustion time scale: star
formation is active despite the $HI$ deficit, which is indicative
of a high star formation efficiency (Zasov 1987; Scodeggio and
Gavazzi 1993; Kennicutt et al. 1984).

It is worth noting that the star formation rates are statistically
correlated with the volume gas density: $SFR\sim \rho_{gas}^n$ where
$n\approx 1-2$ (Schmidt's law; see Abramova and Zasov, 2008 for a
discussion). Therefore, even a twofold increase in gas density
accelerates the gas spending on star formation by a factor of $2-4$.
For spiral galaxies, the gas exhaustion time scale $T_g=
M_{gas}/SFR$ is usually several Gyr (Kennicutt 1998, Wong and Blitz
2002; Zasov and Abramova 2006). The dynamic time scale in which the
galaxy crosses the densest inner cluster region is of the same order
of magnitude. Therefore, an increase in gas density and the
corresponding decrease in $T_g $ as the gas layer is compressed can
be an important factor in the evolution of the gas content for
cluster galaxies. A decrease in gas density in the outer disk
regions should facilitate the sweeping-up of the remnants of the
interstellar medium by the flow of intergalactic gas.

In combination with the dynamic pressure and with the minor merging
process of galaxies, the static pressure allows us to explain the
existence of a large number of disk galaxies with low interstellar
gas content (S0- galaxies)in the inner regions of clusters.

(2) The $HI$-deficient galaxies in clusters are distinguished by a
higher (on average) content of the molecular gas with respect to the
atomic one (Kenney and Young 1988, 1989); this cannot always be
explained by the sweeping-up of HI from the peripheral regions of
galaxies and a high $H_2$ concentration toward the center. Indeed,
the example of spiral galaxies in the inner region of the Virgo
cluster shows that an unusually high fraction of the molecular gas
is observed even in the inner disk region, where $HI$ is retained,
although it can be swept up from the disk periphery (Nakanishi et
al. 2006). In the above paper, it was suggested that the external
pressure plays a possible role in increasing the amount of molecular
gas. Gas compression should actually contribute to the
transformation of the atomic gas into the molecular one, since, as
analysis of the available observational data shows, the content of
the latter strongly correlates with the pressure in the disk
midplane (see Kasparova and Zasov 2008; Blitz and Rosolowski 2006;
and references therein). However, this effect is so far difficult to
test quantitatively.

(3) The example of $HI$-deficient galaxies in the Virgo cluster
shows that, in some cases, the decrease in the total amount of
atomic gas in the galaxy is related not so much to a reduction of
the region occupied by it, which is natural to associate with the
ram pressure, as to a decrease in $HI$ surface density over the
entire disk (Cayatte et al. 1994). The latter can be the result of
gas compression and its faster ``exhaustion''.

(4) The magnetic field enhancement  expected when the gas disk in a
galaxy is compressed agrees well with the fact that, as was noted by
several authors, a significant fraction of the spiral galaxies in
clusters are distinguished by a higher intensity of the synchrotron
radiation coming from the disk than that for galaxies outside
clusters (see Scodeggio and Gavazzi (1993), Reddy and Yin (2004) and
references therein).

(5) The role of the static pressure should be most significant for
the galaxies located in the inner region of a cluster filled with an
X-ray-emitting gas and having low velocities with respect to it,
which takes place if the galaxies do not go far away from the
cluster center. In this respect, the $HI$ observations in the
Pegasus I cluster deserve a special attention. A small, but
confidently detectable $HI$ deficit was found for the galaxies in
the central part of this cluster (Levy et al. 2007). Since they have
a very low velocity dispersion, the value of $n_e V_c^2$,
characterizing the dynamic pressure, is more than two orders of
magnitude lower than that in the Coma or Virgo cluster. As a result,
the $HI$ deficit cannot be associated with the motion of the
galaxies in a gas medium in a standard way. The mean electron number
density $<n_e>$ in a cluster, calculated for the model of a
homogeneous sphere, filled with a hot gas with $T = (0.6-3)\cdot
10^7$K is about $2\cdot 10^{-4}$cm$^{-3}$ (Canizares et al. 1986).
This value is almost equal to the lower limit for the particle
number density within 16 arcmin (360 kpc) of the central galaxy of
the cluster, NGC 7619, obtained from ROSAT measurements (Trinchieri
et al. 1997). At $T\approx 1-2$keV, the corresponding gas pressure
is $P\approx (4-8)\cdot 10^3$K$\cdot$cm$^{-3}$. This static
pressure, though it is low, is nevertheless comparable to the
interstellar gas pressure in the outer regions of galactic disks
(see the Introduction); it exceeds the dynamic pressure on the
galaxies in the central part of the cluster, which, based on the
estimate $n_e V^2 =12(km/s)^2cm^{-3}$ (Levy et al. 2007), is $\simeq
1.5\cdot 10^3$K$\cdot$cm$^{-3}$, by a factor of several. Therefore,
in this case, the static pressure of the gas can be more
significant.

The external static pressure on the gas disk of a galaxy can be
exerted not only by the intergalactic gas, but also by the gas
inside the galaxy located in the bulge or inner halo, if the gas has
a fairly high density, $\sim 10^{-2} - 10^{-3}$cm$\!^{-3}$, at a
virial temperature of several million degrees. The available
estimates of the hot-gas density and temperature in the bulges and
halos of galaxies based on their soft X-ray emission are so far few
in number. However, they show that a medium with the required
density and temperature actually exists at least in some of the
massive galaxies, such as Ml04, NGC 4565, NGC 5746, NGC 4921, and
NGC 4911 (Wang 2006; Yao and Wang 2007; Rasmussen et al. 2006; Sun
et al. 2007). In these cases, the hot-gas pressure on the disk
manifests itself in the same way as the pressure of the
intergalactic medium: it compresses the gas in the inner disk region
and increases its volume density and, as a sequence, the mass
fraction of the molecular gas, which affects the star formation.
Indeed, observations show that in the inner regions of such galaxies
as M81, M106, and, possibly, M31, where the stellar bulge dominates,
the fraction of the molecular gas is considerably higher than could
be expected from the semi-empirical dependence of the relative mass
of the molecular gas on the gas pressure, if the latter is estimated
without taking into account the external medium (Kasparova and Zasov
2008).

Note that at an earlier evolutionary stage of  galaxies, their gas
disk had a much higher surface density, which then decreased when
the bulk of the gas passed into stars. For this reason, the
efficiency of the gas sweeping-up by the ram pressure was lower than
that at present. Nevertheless, if the pressure of surrounding gas
was the same as in the present time, the compression of the gas
layer could accelerate the star formation and play an important role
in forming the outer regions of stellar disks.

Thus, under some quite realistic conditions, the pressure of the hot
medium on the gas disk of a galaxy can be an important factor of its
evolution.

\bigskip
\section{ACKNOWLEDGMENTS}

This work was supported by the Russian Foundation for Basic
Research (project nos. 07-02-00792 and 07-02-01204).

\section{REFERENCES}
\bigskip

\small [1] O. V. Abramova and A. V. Zasov, Astron. Rep., 52, 257
(2008), arXiv:0712.1149 (2008).

 \noindent [2] M. Arnaud and A. E. Evrard, Mon. Not. R. Astron.
Soc. 305, 631 (1999).

 \noindent [3] L. Blitz and E. Rosolowsky, Astrophys. J. 650, 933
(2006).

 \noindent [4]  C. R. Canizares, M. N. Donahue, G. Trinchieri, et al.,
Astrophys. J. 304, 312 (1986).

 \noindent [5] V. Cayatte, C. Kotanyi, C. Balkowski, and van
J. H. Gorkom, Astron. J. 107, 1003 (1994).

 \noindent [6] D. P. Cox, Ann. Rev. Astrophys. Astron. 43, 337
(2005).

  \noindent [7] M. Dahlem and I. Thiering, Publ. Astron. Soc. Pacific
112, 158 (2000).

   \noindent [8] A. Finoguenov, T. H. Reiprich, and H. Boehringer,
Astron. Astrophys. 368, 749 (2001).

 \noindent [9] A. V. Kasparova and A. V. Zasov, Astron. Lett, 34, 152 (2008),
 arXiv:0802.3804 (2008).

\noindent [10] J. D. P. Kenney and J. S. Young, Astrophys. J. 66,
261 (1988).

\noindent [11] J. D. P. Kenney and J. S. Young, Astrophys. J. 344,
171 (1989).

\noindent [12] R. C. Kennicutt, Astrophys. J. 498, 541 (1998).

\noindent [13] R. C. Kennicutt, G. D. Bothun, and R. A. Schommer,
Astron. J. 89, 1279 (1984).

\noindent [14] T. Kronberger, W. Kapferer, C. Ferrari, Astron.
Astrophys. 481, 337 (2008).

\noindent [15] L. Levy, J. A. Rose, I . H. van Gorkom, and B.
Chaboyer, Astron. J. 133, 1104 (2007).

\noindent [16] H. Nakanishi, N. Kuno, Y. Sofue, et al., Asrophys. J.
651, 804 (2006).

\noindent [17] J. Nevalainen, R. Lieu, M. Bonamente, and D. Lumb,
Astrophys. J. 584, 716 (2003).

\noindent [18] P. E. J. Nulsen and H. Bohringer, Mon. Not. R.
Astron. Soc. 274, 1093 (1995).

\noindent [19] J. Rasmussen, J. Sommer-Larsen,K. Pedersen, et al.,
atro-ph/0610893 (2006).

\noindent [20] N.A. Reddy, M.S. Yin,  Astrophys. J., 600, 695,
(2004).

\noindent [21] M. Scodeggio and G. Gavazzi, Astrophys. J. 409, 110
(1993).

\noindent [22] C. Sengupta, R. Balasubramanyam, and K.
Dwarakanath,Mon. Not. R. Astron. Soc. 378, 137 (2007).

\noindent [23] M. Sun, C. Jones, W. Forman, et al., Astrophys. J.
657, 197 (2007).

\noindent [24] G. Trinchieri, G. Fabbiano, and D.-W. Kim, Astron.
Astrophys. 318, 361 (1997).

\noindent [25] A. Vikhlinin, M. Markevitch, W. Forman and C. Jones,
Astrophys. J. 555, L87 (2001).

\noindent [26] Q. D.Wang, astro-ph/0611038 (2006).

\noindent [27] T.Wong and L. Blitz, Astrophys. J. 569, 157 (2002).

\noindent [28] Y. Yao and Q. D.Wang, astro-ph/0705.2772 (2007).

\noindent [29] A. V. Zasov, Pis'ma Astron. Zh. 13, 757 (1987) [Sov.
Astron. Lett. 13, 319 (1987)].

\noindent [30] A. V. Zasov and O. V. Abramova, Astron. Zh. 83, 976
(2006) [Astron. Rep. 50, 874 (2006)].

\end{document}